\documentclass[twocolumn,times]{aastex631}
\usepackage{amsmath}

\shorttitle{Model for Abundances in Metal-poor Stars}
\shortauthors{Gross et al.}

\begin{document}

\title{A Data-Driven Model for Abundances in Metal-poor Stars and Implications for Nucleosynthetic Sources}
\author[0000-0002-7893-4183]{Axel Gross}
\affiliation{School of Physics and Astronomy,
      University of Minnesota, Minneapolis, MN 55455, USA} 
\author[0000-0002-2385-6771]{Zewei Xiong}
\affiliation{GSI Helmholtzzentrum f{\"u}r Schwerionenforschung, Planckstraße 1, D-64291 Darmstadt, Germany} 
\author[0000-0002-3146-2668]{Yong-Zhong Qian}
\affiliation{School of Physics and Astronomy,
      University of Minnesota, Minneapolis, MN 55455, USA}

\begin{abstract}
We present a data-driven model for abundances of Fe, Sr, Ba, and Eu in metal-poor (MP) stars.
The production patterns for core-collapse supernovae (CCSNe) and binary neutron star mergers (BNSMs)
are derived from the data of \cite{holmbeck2020} on [Sr/Fe], [Ba/Fe], and [Eu/Fe] for 195 stars.
Nearly all the data can be accounted for by mixtures of contributions from these two sources.
We find that on average, the Sr contribution to an MP star from BNSMs is $\approx 3$ times that from CCSNe. 
Our model is also consistent with the solar inventory of Fe, Sr, Ba, and Eu. We carry out a parametric
$r$-process study to explore the conditions that can give rise to our inferred production patterns and find that such
conditions are largely consistent with those from simulations of CCSNe and BNSMs. Our model 
can be greatly enhanced by accurate abundances of many $r$-process elements in a large number of MP stars, 
and future results from this approach can be used to probe the conditions in CCSNe and BNSMs in much more detail.
\end{abstract}


%%%%%%%%%%%%%%%%%%%%%%%%%%%%%%%%%%%%%%%%%%%%%%%%%%%%%%%%%%%%%%%%%%%%%%%%%%%%%%%%
\section{Introduction} \label{sec:intro}

It is well known that Type Ia (SNe Ia) and core-collapse supernovae (CCSNe) are major sources for Fe, 
that elements heavier than the Fe group are mainly produced by the rapid ($r$) and slow ($s$) neutron-capture processes,
and that asymptotic giant branch (AGB) stars of low to intermediate masses are the site of the main $s$-process producing Sr 
and heavier elements (see e.g., \citealt{arcones2023} for a review). 
The spectacular multimessenger observations of GW170817 \citep{abbott2017} provided strong support of binary neutron star mergers 
(BNSMs) being a site of the $r$-process (e.g. \citealt{kasen2017}), and many theoretical studies have been devoted to this topic both before 
(see e.g., \citealt{thielemann2017} for a review) and after this event (e.g., \citealt{curtis2023,just2023,kiuchi2023}).
In addition, theoretical studies suggest that CCSNe may produce some elements heavier than the Fe group 
(e.g., \citealt{woosley1992,hoffman1997,wanajo2018,wang2023}) and 
that a subset of them may even be a site of the $r$-process (e.g., \citealt{nishimura2015,siegel2019,fischer2020}). 

Despite the above advances, we are still far from being able to make 
precise ab initio predictions for the nucleosynthesis of astrophysical sources. In particular, the extreme conditions 
in the dynamic environments of CCSNe and BNSMs are inherently difficult to simulate, and there are large uncertainties in the nuclear 
input for simulating these sources and the associated $r$-process.
On the other hand, because both sources are associated with rapidly-evolving massive stars, they are expected to have dominated the chemical evolution of
the universe during the first $\sim 1$~Gyr, before Fe contributions from SNe Ia and $s$-process contributions from AGB stars became significant.
Consequently, metal-poor (MP) stars formed during this early epoch provide an excellent fossil record for deciphering the nucleosynthesis of CCSNe and BNSMs.
For example, \cite{qian2001,qian2008} took the observed elemental abundance patterns in two MP stars as the production patterns of two distinct sources
and showed that the data for other stars could be largely explained as mixtures of those two patterns. With hindsight, they should have identified
those two sources as CCSNe and BNSMs rather than two distinct subsets of CCSNe.

In this Letter we take a data-driven approach to infer the average production patterns of CCSNe and BNSMs. 
Unlike \cite{qian2001,qian2008}, we do not take these patterns from two individual MP stars.
Instead, we derive them from the latest data on [Sr/Fe], [Ba/Fe], and [Eu/Fe] \citep{holmbeck2020}
provided by the $R$-Process Alliance (RPA) search for $r$-process-enhanced 
stars in the Galactic Halo. We attribute the pattern with dominant production of Fe and Sr to CCSNe and 
that with dominant production of Sr, Ba, and Eu to BNSMs. We show that nearly all the RPA data can be accounted for by mixtures of contributions from these two sources, and that their contributions over the Galactic history are also consistent with the solar inventory of Fe, Sr, Ba, and Eu (\S\ref{sec:model}).
We then carry out a parametric study of the $r$-process to explore the conditions that can give rise to the inferred production patterns 
and compare such conditions with those found in simulations of CCSNe and BNSMs (\S\ref{sec:para}). 
Finally, we summarize our results and discuss how our approach can be greatly enhanced by accurate abundances of many $r$-process elements 
in a large number of MP stars and how future results from this approach can be used to further probe the conditions in CCSNe and BNSMs (\S\ref{sec:dis}).

\section{Model for Abundances in MP Stars}
\label{sec:model}

We model the abundance of element E in an MP star as a mixture of contributions from two distinct sources, each with a fixed 
characteristic production pattern. The number ratio of E to Fe atoms in the star is given by
\begin{equation}
    \left(\frac{\rm E}{\rm Fe}\right)=x\left(\frac{\rm E}{\rm Fe}\right)_1+(1-x)\left(\frac{\rm E}{\rm Fe}\right)_2,
    \label{eq:efe}
\end{equation}
where $({\rm E/Fe})_1$ and $({\rm E/Fe})_2$ represent the production of E relative to Fe by sources 1 and 2, respectively, 
and $x$ is the fraction of Fe contributed by source 1. Because abundance data are commonly presented in terms of
$[{\rm E/Fe}]=\log({\rm E/Fe})-\log({\rm E/Fe})_\odot$, we rewrite Eq.~(\ref{eq:efe}) as
\begin{equation}
    10^{[{\rm E/Fe}]}=x\times 10^{[{\rm E/Fe}]_1}+(1-x)\times 10^{[{\rm E/Fe}]_2}.
    \label{eq:efe2}
\end{equation}

The RPA data \citep{holmbeck2020} consist of complete measurements of [Sr/Fe], [Ba/Fe], and [Eu/Fe] for 211 MP stars with 
$-3\lesssim[{\rm Fe/H}]\lesssim-1$. We suspect that 16 of these stars received $s$-process contributions (from relatively
massive AGB stars or through binary mass transfer from former AGB companions) based on their high values of $[{\rm Ba/Eu}]>0$, and therefore,
exclude them from further consideration. As a first test of the model, we 
determine two sets of parameters $\{[{\rm Sr/Fe}]_i,[{\rm Ba/Fe}]_i, [{\rm Eu/Fe}]_i\}$ $(i=1,2)$ that best reproduce all the data for the 
remaining 195 stars taking into account the uncertainty of $\sigma\approx 0.2$~dex for each measurement. We obtain these parameters
by minimizing
\begin{equation}
 Q=\sum_{j,{\rm E}} H\left(\frac{[{\rm E/Fe}]_j-[{\rm E/Fe}]_{*,j}}{\sigma}\right),
 \label{eq:q}
\end{equation}
where $[{\rm E/Fe}]_j$ and $[{\rm E/Fe}]_{*,j}$ refer to the predicted value and the measured mean for the $j$th star, respectively, 
and $H(y)$ is the Huber loss function defined by $H(y)=y^2/2$ for $|y|\leq 1$ and $H(y)=|y|-1/2$ for $|y|>1$. 
The use of $H(y)$ reduces sensitivity to the outliers in the data. With $\{[{\rm Sr/Fe}]_1,[{\rm Ba/Fe}]_1, [{\rm Eu/Fe}]_1\}=\{-0.49,-3.00,-0.77\}$
and $\{[{\rm Sr/Fe}]_2,[{\rm Ba/Fe}]_2, [{\rm Eu/Fe}]_2\}=\{0.90,0.91,1.30\}$, we find that
all the data on [Sr/Fe], [Ba/Fe], and [Eu/Fe] for 140 (190) out of the 195 stars can be reproduced within $1\sigma$ ($2\sigma$).
Note that the fraction $x$ of Fe contributed by source 1 to each star is also optimized during the above calculation.

The above production patterns are inferred from a mathematical procedure without considering the mechanisms of synthesizing Fe, Sr, Ba, and Eu.
In this sense, they represent important observational constraints on the sources for these elements. Sources 1 and 2 have drastically different 
production of Sr, Ba, and Eu relative to Fe. With respect to the solar composition, source 1 has high Fe production while source 2 has high production
of Sr, Ba, and Eu. The extremely low production of Ba by source 1 with $[{\rm Ba/Fe}]_1=-3.00$ is especially striking. Because the above characteristics 
of sources 1 and 2 bear strong resemblance to those of CCSNe and BNSMs, respectively, we carry out a second test of our model by incorporating
predefined features of these two sources. Specifically, we assume that CCSNe produce Fe and Sr but no Ba or Eu while BNSMs produce Sr, Ba, and Eu
but no Fe. In this simplified but well-motivated case, the production patterns are characterized by $[{\rm Sr/Fe}]_{\rm SN}$, $[{\rm Ba/Sr}]_{\rm NSM}$, and $[{\rm Eu/Ba}]_{\rm NSM}$. 
The values of [Sr/Fe], [Ba/Fe], and [Eu/Fe] for a star are given by
\begin{eqnarray}
{\rm [Sr/Fe]}&=&{\rm [Sr/Fe]}_{\rm SN}+\log(1+\alpha),\label{eq:srfe}\\
{\rm [Ba/Fe]}&=&{\rm [Sr/Fe]}_{\rm SN}+{\rm [Ba/Sr]}_{\rm NSM}+\log\alpha,\label{eq:bafe}\\
{\rm [Eu/Fe]}&=&{\rm [Ba/Fe]}+{\rm [Eu/Ba]}_{\rm NSM},\label{eq:eufe}
\end{eqnarray}
where $\alpha$ is the ratio of the Sr contribution from BNSMs to that from CCSNe. Note that $\alpha$ plays a similar role to $x$ in Eq.~(\ref{eq:efe2})
because now only Sr is produced by both sources. By minimizing $Q$ in Eq.~(\ref{eq:q}), we find that
with $[{\rm Sr/Fe}]_{\rm SN}=-0.54$, $[{\rm Ba/Sr}]_{\rm NSM}=0.00$, and $[{\rm Eu/Ba}]_{\rm NSM}=0.43$, all the data on [Sr/Fe], [Ba/Fe], and [Eu/Fe] for
141 (189) out of the 195 stars can be reproduced within $1\sigma$ ($2\sigma$). So the same level of agreement with the data is achieved for 
the model represented by Eq.~(\ref{eq:efe2}) and that by Eqs.~(\ref{eq:srfe})--(\ref{eq:eufe}). We focus on the latter in the discussion below.

\begin{figure}
    \epsscale{1.1}
    \plotone{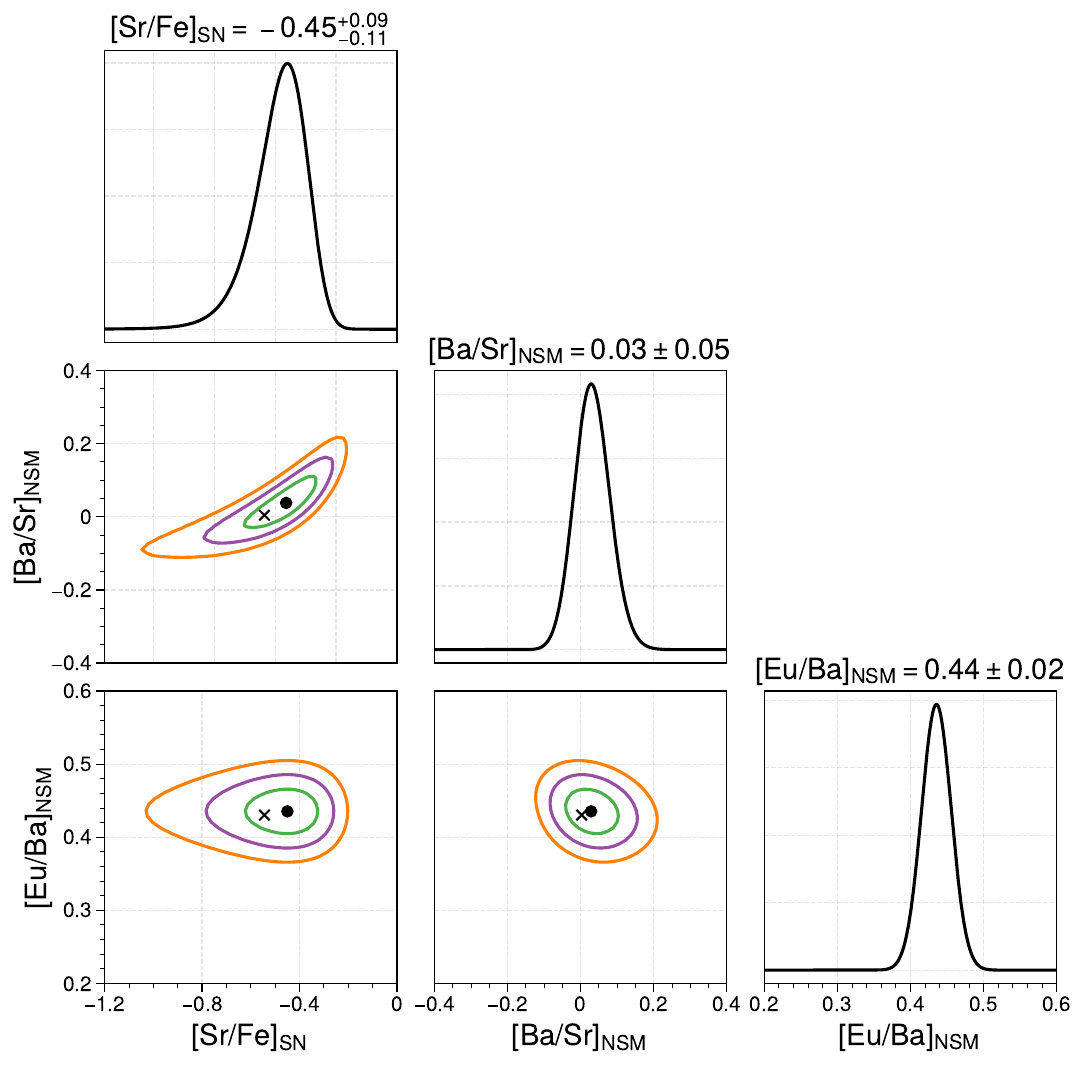}
    \caption{Corner plot for parameters characterizing the production patterns of CCSNe and BNSMs. 
    The marginal distribution of each parameter as well as $1\sigma$, $2\sigma$, and $3\sigma$ contours for pairs of parameters are shown.
    Filled circles indicate the best-fit parameters, which are close to those (crosses) inferred from a different method.}
    \label{fig:corner}
\end{figure}

Next, we estimate the uncertainties in the inferred production patterns using Bayesian techniques.
For the $j$th star, the likelihood of reproducing the data $D_j=\{{\rm [Sr/Fe]}_{*,j},{\rm [Ba/Fe]}_{*,j},{\rm [Eu/Fe]}_{*,j}\}$ by the model 
with the parameters $M=\{{\rm [Sr/Fe]}_{\rm SN},{\rm [Ba/Sr]}_{\rm NSM},{\rm [Eu/Ba]}_{\rm NSM}\}$ and $A_j=\log\alpha_j$ is
\begin{equation}
    P_j(D_j|M,A_j)={\cal N}_j({\rm Sr}){\cal N}_j({\rm Ba}){\cal N}_j({\rm Eu}),
\end{equation}
where, for example, ${\cal N}_j({\rm Sr})$ is the normal distribution of ${\rm [Sr/Fe]}_j$ centered at ${\rm [Sr/Fe]}_{*,j}$ 
with a standard deviation of $\sigma$. Assuming uniform prior probabilities for $M$ and all the $A_j$'s, we obtain the posterior probability of the model
\begin{equation}
P(M,\{A_j\}|\{D_j\})=\frac{\prod_jP_j(D_j|M,A_j)}{P(\{D_j\})},
\end{equation}
where $P(\{D_j\})={\int dM\prod_j\int dA_jP_j(D_j|M,A_j)}$.
Various marginal distributions can be obtained by integrating $P(M,\{A_j\}|\{D_j\})$ over the parameters of no concern. For example,
$P({\rm [Sr/Fe]}_{\rm SN})$, $P({\rm [Sr/Fe]}_{\rm SN}$, $P({\rm [Ba/Sr]}_{\rm NSM})$, and similar distributions are presented in Fig.~\ref{fig:corner}.
We find ${\rm [Sr/Fe]}_{\rm SN}=-0.45^{+0.09}_{-0.11}$, ${\rm [Ba/Sr]}_{\rm NSM}=0.03\pm0.05$, and
${\rm [Eu/Ba]}_{\rm NSM}=0.44\pm0.02$. The optimal values found above by the second test of the model are in good agreement with these results 
(within $1\sigma$).

\begin{figure}
    \epsscale{1.1}
    \plotone{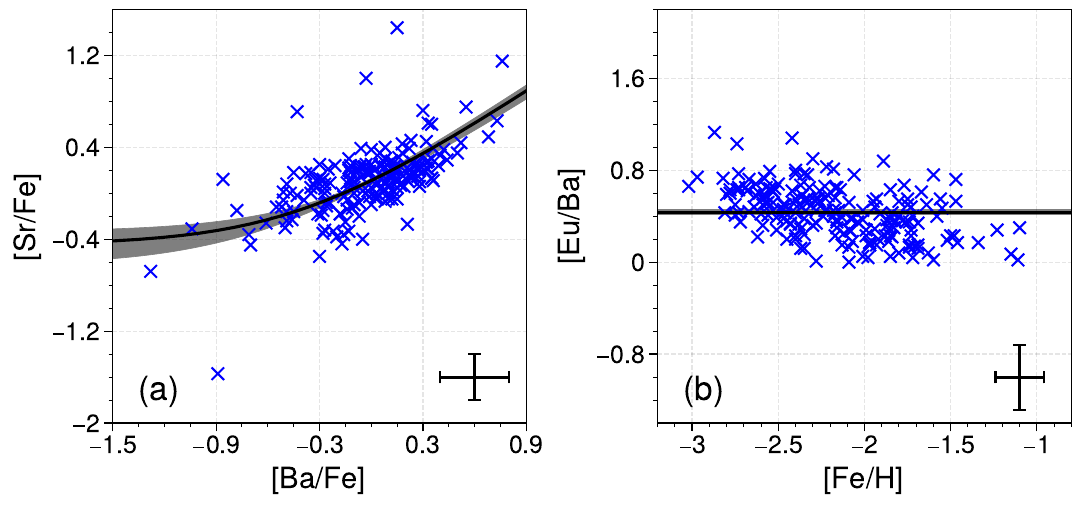}
    \caption{Comparison of the model with the data (crosses) for (a) [Sr/Fe] and [Ba/Fe], and (b) [Eu/Ba] and [Fe/H].
    The dark curve corresponds to the best-fit parameters and the gray region reflects $1\sigma$ uncertainties in the parameters.
    The error bars indicate the uncertainties in each measurement.}
    \label{fig:comp}
\end{figure}

Because CCSNe are the only source for Fe and BNSMs are the only source for Ba and Eu, the model predicts
\begin{eqnarray}
10^{[{\rm Sr/Fe}]}&=&10^{[{\rm Sr/Fe}]_{\rm SN}}+10^{[{\rm Ba/Fe}]-[{\rm Ba/Sr}]_{\rm NSM}},\\
{\rm [Eu/Ba]}&=&[{\rm Eu/Ba}]_{\rm NSM}.
\end{eqnarray}
The above predictions are compared with the data in Fig.~\ref{fig:comp}.
Taking into account the measurement errors, we find that nearly all the stars follow these predictions. 
The same is also true of the relation between [Sr/Fe] and [Eu/Fe] (not shown), which follows from the above two predictions.
In Fig.~\ref{fig:comp}a, one star lies far below while three lie significantly above the relation between 
[Sr/Fe] and [Ba/Fe]. We will discuss these anomalous stars further in \S\ref{sec:dis}.

\begin{figure}
    \epsscale{1.1}
    \plotone{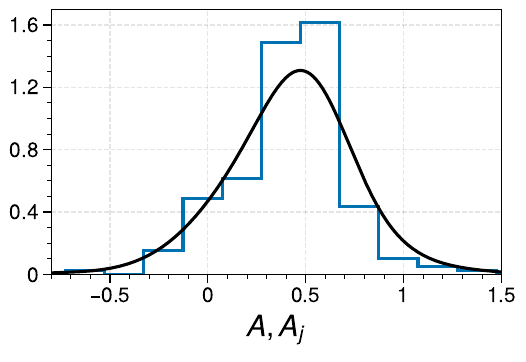}
    \caption{Normalized histogram of optimal $A_j$ for each star and distribution of $A$ for MP stars. The latter (black curve) is taken to be
    the algebraic mean of all the marginal distributions of $A_j$.}
    \label{fig:mix}
\end{figure}

The normalized histogram of the optimal $A_j$ for each star is shown in Fig.~\ref{fig:mix}. We take the algebraic mean of 
all the marginal distributions of $A_j$ to be the distribution of $A=\log\alpha$ for MP stars, which is also shown in Fig.~\ref{fig:mix}.
This distribution is very similar to the histogram and gives $A=0.47^{+0.29}_{-0.35}$, which means that on average,
the Sr contribution to a star from BNSMs is $\alpha=10^A\approx 3$ times that from CCSNe. Assuming that CCSNe and BNSMs have 
operated the same way over the Galactic history, we expect that the solar system material represents the average mixture of their 
contributions very well, and therefore,
\begin{equation}
    \alpha_\odot\approx\frac{10^{{\rm [Sr/Eu]}_{\rm NSM}}}{10^{{\rm [Sr/Fe]}_{\rm SN}}}
    \frac{{\rm (Fe/H)}_\odot}{{\rm (Fe/H)}_{\odot,{\rm SN}}}\approx 3,
\end{equation}
where ${\rm (Fe/H)}_{\odot,{\rm SN}}$ is the CCSN contribution to the solar Fe inventory, and we have taken ${\rm (Eu/H)}_{\odot,{\rm NSM}}\approx{\rm (Eu/H)}_\odot$ because almost all of the solar Eu inventory came from
the $r$-process. With ${\rm [Sr/Eu]}_{\rm NSM}=-{\rm [Ba/Sr]}_{\rm NSM}-{\rm [Eu/Ba]}_{\rm NSM}=-0.47$ and ${\rm [Sr/Fe]}_{\rm SN}=-0.45$,
we obtain ${\rm (Fe/H)}_{\odot,{\rm SN}}\approx {\rm (Fe/H)}_\odot/3$. Studies of Galactic chemical evolution 
(e.g., \citealt{matt1986,timmes1995}) estimated that CCSNe contributed $\approx 1/3$ to $2/3$ of the solar Fe inventory.
In view of the very different approaches taken here and in those studies, it is remarkable that they give similar results.

Our model predicts that the net $r$-process contribution from CCSNe and BNSMs to the solar Sr inventory is
\begin{equation}
    \frac{{\rm (Sr/H)}_{\odot,r}}{{\rm (Sr/H)}_\odot}\approx\frac{1+\alpha_\odot}{\alpha_\odot}\times 10^{{\rm [Sr/Eu]}_{\rm NSM}}
    \approx0.45,
\end{equation}
and that the net $r$-process contribution from BNSMs to the solar Ba inventory is
\begin{equation}
    \frac{{\rm (Ba/H)}_{\odot,r}}{{\rm (Ba/H)}_\odot}\approx 10^{-{\rm [Eu/Ba]}_{\rm NSM}}
    \approx0.36.
\end{equation}
The conventional way of estimating the solar $r$-process inventory of an element was to subtract the $s$-process contribution
from its net solar inventory. Because the $s$-process contribution was estimated from a parametric model, this procedure could 
have large uncertainties. For example, for $^{88}$Sr, the dominant isotope of Sr, \cite{goriely1999} estimated that the $r$-process 
contributed $\approx 23\%$ of its solar inventory, but with a possible range of (0--27)\%. While the higher end of this range
is $\approx 1.6$ times lower than our estimate, the discrepancy is less in terms of the $s$-process contribution,
with his estimate (73\%) being $\approx 1.3$ times higher than ours (55\%).
\cite{goriely1999} also estimated that the $r$-process contributed $\approx 15\%$ of the solar inventory of $^{135,137,138}$Ba, the
dominant isotopes of Ba, but with a possible range of (0--38)\%. Our estimate is just below the higher end of this range.
Considering the drastically different approach used here, we regard our estimates of the $r$-process contributions
to the solar inventory of Sr and Ba as quite reasonable. \cite{qian2001} advocated similar estimates
to make their model fit the data on MP stars available then. The difference is that our estimates are
the direct consequences of our model while theirs were introduced into their model as corrections.

\section{Implications for CCSNe and BNSMs}
\label{sec:para}
Because CCSNe and BNSMs are inherently difficult to simulate and there are large uncertainties 
in the nuclear input for simulating these sources and the associated $r$-process, it is not very instructive
to compare the production patterns calculated from specific simulations with the average patterns derived above from the data on MP stars.
Instead, we carry out a parametric study of nucleosynthesis to explore the astrophysical conditions that may produce 
our inferred patterns, and then compare such conditions with those found in simulations.
In each parametric run, we follow the expansion of some ejecta that could occur in CCSNe or BNSMs.
We start each run at time $t=0$ from an initial state with temperature $T_0=10$~GK, density $\rho_0$, 
and electron fraction $Y_{e,0}$. The subsequent evolution of density is specified by
\begin{equation}
    \rho(t) = \begin{cases}
        \rho_0 e^{-t/\tau}, & t\leq t_{\rm tr}, \\
        \rho_0 e^{-t_{\rm tr}/\tau}(t_{\rm tr}/t)^3, & t>t_{\rm tr},
    \end{cases}
\end{equation}
where $\tau$ is a characteristic timescale and $t_{\rm tr}=3(1-\ln 0.6)\tau$ corresponds to 
the transition between the two expansion regimes as suggested by \cite{lippuner2015}.
We determine $\rho_0$ from $\phi_0=k_B^4 T_0^3m_N/(\hbar^3c^3\rho_0)$,
where $k_B$ is the Boltzmann constant, $\hbar$ is the Planck constant, $c$ is the speed of light, 
and $m_N$ is the nucleon rest mass. We assume nuclear statistical equilibrium between $T_0$ and 
$T=8$~GK and use a reaction network to evolve the nuclear composition for $T<8$~GK. For a specific 
set of $\phi_0$, $\tau$, and $Y_{e,0}$, the change of composition at each time step is 
accompanied by energy release. We calculate the corresponding change of temperature based on the 
thermodynamics associated with this energy release and the expansion as described in \citet{mendoza2015}.
Consequently, $T(t)$, $Y_e(t)$, and the entropy $S(t)$ are updated along with $\rho(t)$ at each time step.

We logarithmically sample $\phi_0$ between 1 and $100\,k_B$ per baryon and
$\tau$ between $10^{-3}$ and 1~s, and uniformly sample $Y_{e,0}$ between 0.05 and 0.55.
Each parameter takes 21 values, so there are a total of 9261 parametric runs.
We use the same nuclear reaction network as employed in \citet{collins2023} and adopt
the nuclear mass model FRDM for the $r$-process (see \citealt{mendoza2015}).
Because the evolution at $T\lesssim 5$~GK is more pertinent to the final nucleosynthesis outcome, we use the expansion timescale 
$\tau_{\rm exp}=|d \ln \rho(t)/dt|^{-1}$ along with the $Y_e$ and $S$ at $T=5$~GK to characterize the outcome of each parametric run below.

\begin{figure}
\epsscale{1.1}
\plotone{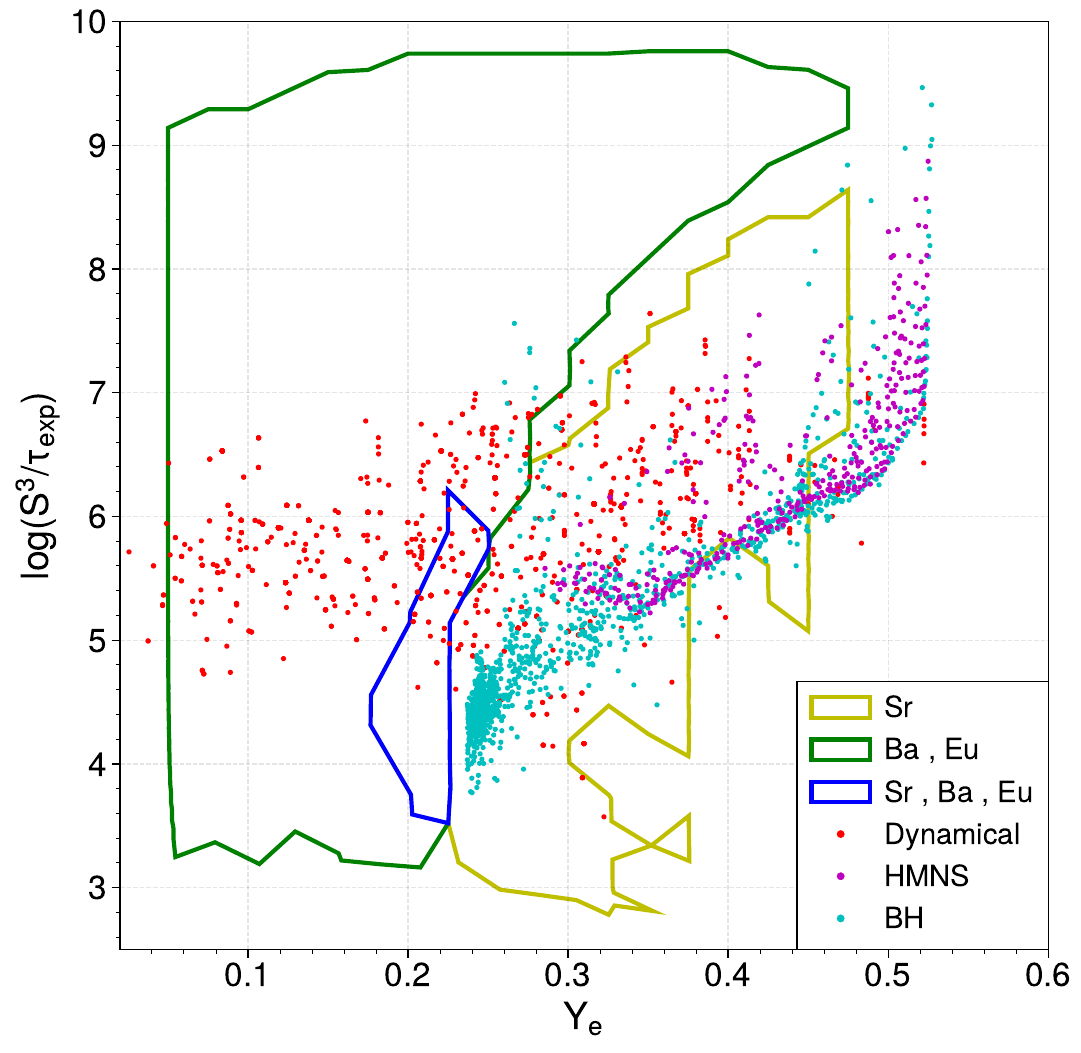}
\caption{Regions of $\log(S^3/\tau_{\rm exp})$ and $Y_e$ for significant production of Sr only, Ba and Eu only, and all of them, respectively.
Conditions in the dynamical ejecta, HMNS wind, and BH torus from the BNSM model sym-n1-a6 simulated by \citet{just2023} are shown for comparison.}
\label{fig:sets}
\end{figure}

Our main interest is in the $r$-process production of Sr, Ba, and Eu, which requires neutron-rich conditions with $Y_e<0.5$.
For fixed $Y_e$, the qualitative outcome of this nucleosynthesis approximately correlates with $S^3/\tau_{\rm exp}$
(e.g., \citealt{hoffman1997}). As shown in Fig.~\ref{fig:sets}, regions of $\log(S^3/\tau_{\rm exp})$ and $Y_e$ fall into three
categories, in which significant production occurs for Sr only, Ba and Eu only, and all of them, respectively. As a quantitative criterion, 
an element is produced significantly if its mass fraction exceeds 10\% of the corresponding value for 
the solar $r$-process pattern.

It was expected long ago that CCSNe are a significant source for Sr \citep[e.g.,][]{woosley1992} with the production 
occurring in the neutrino-driven wind (e.g., \citealt{hoffman1997}). The parameter region with significant production of Sr only
in Fig.~\ref{fig:sets} can be compared to the conditions found in CCSN models
\citep[e.g.,][]{roberts2010,wanajo2018,xiong2019,xiong2020,sieverding2020,wang2023}.
The relevant ejecta have $S\sim 10$--$80\,k_B$ per baryon and $\tau_{\rm exp}\sim0.01$--0.6~s, with lower $S$ associated with 
smaller $\tau_{\rm exp}$. The corresponding range of $\log(S^3/\tau_{\rm exp})$ is $\sim 5$--6, where $S$ is in units of $k_B$ 
per baryon and $\tau_{\rm exp}$ is in units of s. These ejecta typically have $Y_e\sim 0.45$--0.55, but some models have $Y_e$ 
as low as $\sim 0.38$ \citep[e.g.,][]{wanajo2018}, and more models can have similarly low $Y_e$ if $\nu_e$ and $\bar\nu_e$ are
appropriately mixed with sterile species that do not interact with matter \citep[e.g.,][]{xiong2019}. It can be seen from 
Fig.~\ref{fig:sets} that those CCSNe with ejecta having $\log(S^3/\tau_{\rm exp})\sim 5$--6 and $Y_e\sim 0.38$--0.45 would
be a significant source for Sr. With a typical Fe yield of $\sim 0.1\,M_\odot$, CCSNe should produce $\sim 10^{-6}\,M_\odot$ of Sr 
on average to account for ${\rm [Sr/Fe]}_{\rm SN}=-0.45$. This amount is broadly consistent with the mass of the relevant ejecta 
(e.g., \citealt{wang2023}).

BNSMs are favored as the dominant source for $r$-process elements beyond Sr because they have very neutron-rich ejecta.
The various components of their ejecta also facilitate the production of a wide range of $r$-process elements
\citep[e.g.,][]{kiuchi2023,curtis2023,just2023}. The most neutron-rich component is the tidally-disrupted dynamical ejecta, 
which have $Y_e\sim0.05$--0.45, $S\sim 3$--$30\,k_B$ per baryon, and $\tau_{\rm exp}\lesssim 0.01$~s.
Subsequent to the formation of an accretion disk, the ejecta are dominated by the polar neutrino-driven wind during the
life of the hypermassive neutron star (HMNS) and by the outflow, or torus, associated with viscous heating following the collapse of the HMNS
into a black hole (BH). Both the HMNS wind and the BH torus interact with neutrinos, and therefore, have more correlated conditions than 
the dynamical ejecta. The HMNS wind is more affected by neutrinos and has $Y_e\sim 0.3$--0.53, $S\sim 15$--$60\,k_B$ per
baryon, and $\tau_{\rm exp}\sim 4\times 10^{-3}$ to 0.03~s. The BH torus has $Y_e\sim 0.23$--0.45, $S\sim 10$--$30\,k_B$
per baryon, and $\tau_{\rm exp}\sim 0.01$ to 0.3~s. The conditions for each of the above three ejecta components 
from the model sym-n1-a6 simulated by \citet{just2023} are shown in terms of $\log(S^3/\tau_{\rm exp})$ and $Y_e$ in Fig.~\ref{fig:sets}. 
The amount of ejecta in each component as a function of $Y_e$ can
be found in their Fig.~3a. It can be seen from these two figures that both the HMNS wind and
the BH torus mostly have significant production of Sr only, with a very small fraction of the latter having significant production of Ba and Eu only.
In contrast, the dynamical ejecta mostly have significant production of Sr only or Ba and Eu only, with
a small fraction having significant production of Sr, Ba, and Eu. Although the nuclear input for the $r$-process in \citet{just2023}
was different from that in our parametric study,
their results on nucleosynthesis in the three ejecta components are in good qualitative agreement with the above discussion.

\begin{figure}
\epsscale{1.1}
\plotone{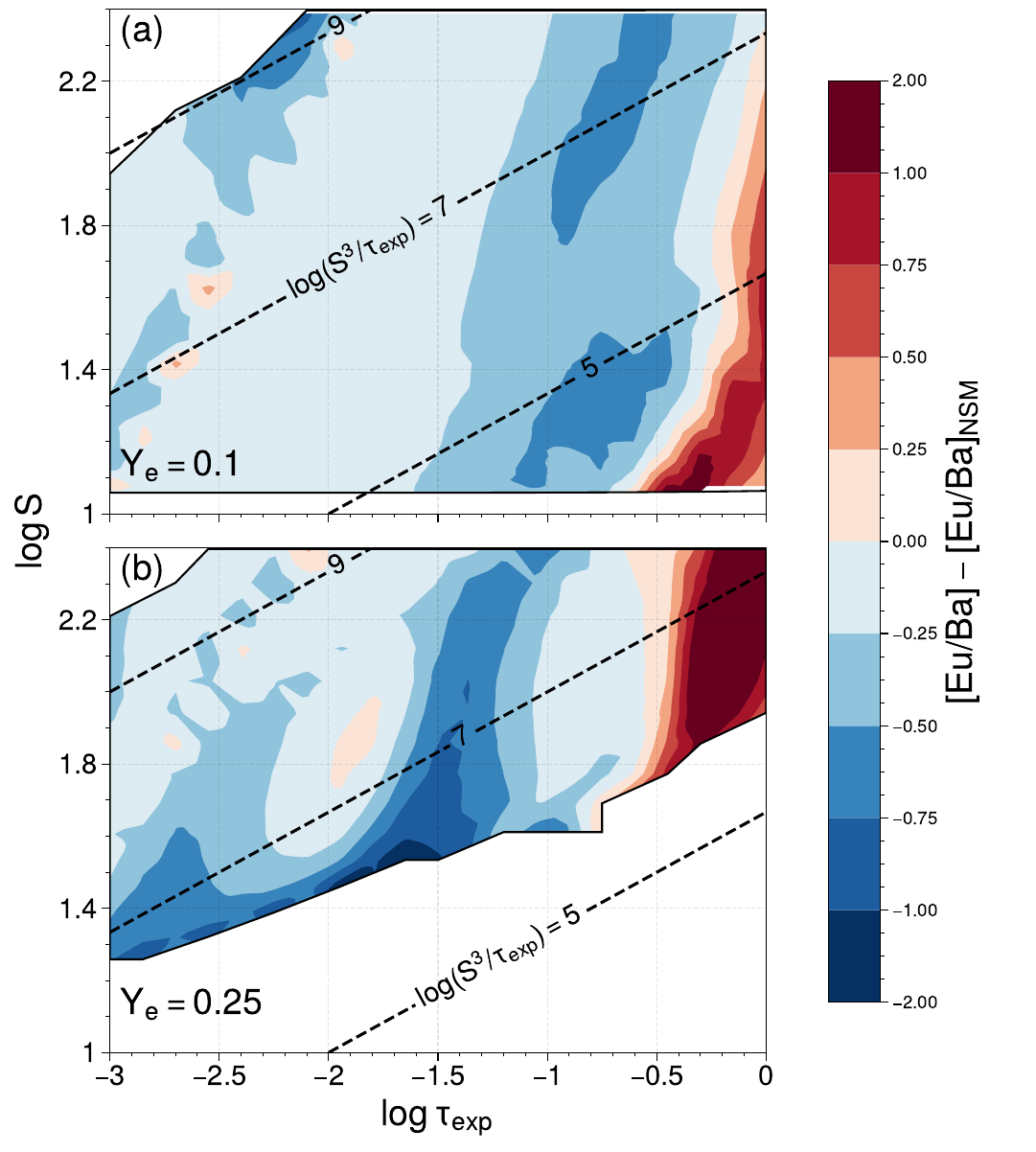}
\caption{Contours of ${\rm [Eu/Ba]}-{\rm [Eu/Ba]}_{\rm NSM}$ as functions of $\log S$ and $\log\tau_{\rm exp}$ for 
(a) $Y_e=0.1$ and (b) $Y_e=0.25$. Solid curves correspond to the boundary for significant production of Ba and Eu only. 
Dashed lines indicate constant values of $\log(S^3/\tau_{\rm exp})$.}
\label{fig:euba}
\end{figure}

We note that while regions of significant production for individual elements are well defined in terms of $S^3/\tau_{\rm exp}$ and $Y_e$, 
the production ratio for a pair of elements is much more sensitive to the detailed conditions, and therefore, is better defined in terms of
$S$, $\tau_{\rm exp}$, and $Y_e$. For example, Fig.~\ref{fig:euba} shows contours of ${\rm [Eu/Ba]}-{\rm [Eu/Ba]}_{\rm NSM}$ as functions of 
$\log S$ and $\log\tau_{\rm exp}$ for $Y_e=0.1$ and 0.25, respectively. It can be seen that for fixed $Y_e$, [Eu/Ba] can change 
drastically for a constant value of $S^3/\tau_{\rm exp}$.

Nonetheless, the regions of significant production in Fig.~\ref{fig:sets} provide good guidance to the conditions that 
could give rise to the inferred BNSM production pattern. In fact, the three categories of regions can be combined in
numerous ways to obtain this pattern. For example, most of the points in the Ba-Eu-only region reproduce the inferred 
${\rm [Eu/Ba]}_{\rm NSM}$ to within a few dexes, so a mixture can easily give the correct result.
Then, such a mixture can be mixed further with any of a large number of points in the Sr-only region to give the inferred 
${\rm [Ba/Sr]}_{\rm NSM}$. The above mixtures can also be obtained with the reasonable requirement that each
type of ejecta involved contribute, for example, at least 10\% of the total mass.
Therefore, we expect that when averaged over various BNSMs, the superposition of the $r$-process production in the dynamical ejecta, 
HMNS wind, and BH torus would account for the inferred BNSM pattern.

\section{Discussion and Conclusions}
\label{sec:dis}
We have presented a data-driven model for abundances of Fe, Sr, Ba, and Eu in MP stars with $-3\lesssim{\rm [Fe/H]}\lesssim -1$.
The production patterns of two distinct sources are derived from the RPA data on [Sr/Fe], [Ba/Fe], and [Eu/Fe] for 195 stars
\citep{holmbeck2020}. We simplify these two sources as CCSNe producing Fe and Sr but no Ba or Eu and BNSMs producing Sr, Ba, 
and Eu but no Fe. Nearly all the data can be accounted for by mixtures of contributions from these two sources, which are 
characterized by ${\rm [Sr/Fe]}_{\rm SN}=-0.45$, ${\rm [Ba/Sr]}_{\rm NSM}=0.03$, and ${\rm [Eu/Ba]}_{\rm NSM}=0.44$.
We find that on average, the Sr contribution to an MP star from BNSMs is $\approx 3$ times that from CCSNe. Assuming that
CCSNe and BNSMs have operated the same way over the Galactic history, we find that CCSNe contributed $\approx 1/3$ of the solar Fe
inventory, in agreement with estimates from studies of Galactic chemical evolution that use Fe yields and rates of occurrence for
CCSNe and SNe Ia. The $r$-process contributions to the solar inventory of Sr and Ba predicted by our model are also reasonable 
in comparison with estimates based on subtraction of the $s$-process contributions when the large uncertainties in the latter 
approach are taken into account.

Four stars show large deviations from the relation between [Sr/Fe] and [Ba/Fe] predicted by our model (see Fig.~\ref{fig:comp}a).
The star J09471921--4127042 with ${\rm [Fe/H]}=-2.67$ has a very low value of ${\rm [Sr/Fe]}=-1.57$ while J15230675--7930072   
with ${\rm [Fe/H]}=-2.55$ has a high value of ${\rm [Sr/Fe]}=0.71$. Their low [Fe/H] values suggest that they might have formed
from materials enriched by a few special CCSNe and BNSMs. On the other hand, for J06195001--5312114 with ${\rm [Fe/H]}=-2.06$
and J06320130--2026538 with ${\rm [Fe/H]}=-1.56$, their high values of ${\rm [Sr/Fe]}=1.00$ and 1.44, respectively, may reflect 
that due to the rarity of BNSMs, deviations from our average BNSM production pattern can still occur up to ${\rm [Fe/H]}\sim -1.6$.
Of course, it is also possible that more than two distinct production patterns are required to characterize CCSNe and BNSMs, even
in the average sense. Unfortunately, with the data on [Sr/Fe], [Ba/Fe], and [Eu/Fe] only, we are unable to derive three well-defined 
production patterns. Clearly, exploration of more than two distinct production patterns requires precise measurements of more 
$r$-process elements in a large number of MP stars.

We have also carried out a parametric study to explore the conditions in CCSNe and BNSMs that may give rise to our inferred production
patterns. We find that such conditions are largely consistent with the results from simulations. We emphasize that
the production ratio for a pair of elements is much more sensitive to the detailed conditions (see Fig.~\ref{fig:euba}).
However, to narrow down the combinations of $Y_e$, $S$, and $\tau_{\rm exp}$ that can give rise to a production pattern,
we need an extensive pattern covering many elements. Therefore, data on abundances of more $r$-process elements are required 
to probe the conditions in CCSNe and BNSMs in more detail. For this purpose and for exploring
the possibility of more than two distinct production patterns, we strongly urge large surveys of MP stars to cover many $r$-process 
elements in addition to Sr, Ba, and Eu.

\section*{Acknowledgments}
We thank Oliver Just for providing the nucleosynthesis conditions from the BNSM model sym-n1-a6. This work was supported in part by
the US Department of Energy under grant DE-FG02-87ER40328 (A.G. and Y.Z.Q.) and by the European Research Council
under ERC Advanced Grant KILONOVA No. 885281 of the European Union’s Horizon 2020 research and innovation program (Z.X.).
The parametric nucleosynthesis calculations were performed on the VIRGO cluster at GSI.
The results from these calculations and the data on MP stars were analyzed with resources of the Minnesota Supercomputing Institute.

\bibliography{ref}
\end{document}